\documentclass[10pt]{article}
\RequirePackage[T1]{fontenc}

\usepackage{comment}
\usepackage{color}
\usepackage{parskip}
\usepackage{appendix}
\usepackage{amssymb}
\usepackage{booktabs}
\usepackage{comment}
\usepackage{graphicx}
\usepackage{caption}
\usepackage{subcaption}
\usepackage{mathrsfs}
\usepackage{tabularx}
\usepackage{arydshln}
\usepackage{amsmath}
\usepackage{multirow}
\usepackage{multicol}
\usepackage{adjustbox}
\usepackage{authblk}

\RequirePackage{graphicx}
\RequirePackage{mathptmx}      
\RequirePackage{flushend}
\RequirePackage[numbers,sort&compress]{natbib}
\RequirePackage[colorlinks,citecolor=blue,urlcolor=blue,linkcolor=blue]{hyperref}

\usepackage[utf8]{inputenc}
\usepackage[T1]{fontenc}
\usepackage{amsmath}
\usepackage{graphicx}
\usepackage[scale=0.8]{geometry}
\usepackage{lineno}
\usepackage{xcolor}
\newcommand{\FCCee}{$\mathrm{FCC_{ee}}\;$}
\newcommand{\FCChh}{$\mathrm{FCC_{hh}}\;$}

\usepackage{sectsty}

\sectionfont{\fontsize{13}{14}\selectfont}
\subsectionfont{\fontsize{11}{12}\selectfont}

\makeatletter
\renewcommand\@author{
    \AB@authlist\\[\affilsep]
    \begin{multicols}{2}
      \begin{quotation}
        \AB@affillist
      \end{quotation}
    \end{multicols}
    }
\def\@maketitle{
  \begin{center}
  \let \footnote \thanks
    {\Large\bfseries\@title}
    {\normalsize
      \begin{center}%
        \baselineskip=12pt
        \@author
      \end{center}\par}
    {\large \@date}
  \end{center}%
}
\makeatother

\title{\Large\textbf{French HEP community input to the European Strategy for Particle
    Physics}}

\author[1]{\normalsize \vspace{0.3cm} \textit{This document sumarizes the French community input to ESPP 2026 and was edited by: }\\ \vspace{0.2cm} Yasmine Amhis}
\author[2]{Jeremy Andrea}
\author[1]{Etienne Aug\'e}
\author[3]{Sara Bolognesi}
\author[3]{Maarten Boonekamp}
\author[4]{Samuel Calvet}
\author[3]{Emilien Chapon}
\author[5]{Didier Contardo}
\author[3]{Fabrice Couderc}
\author[6]{Sabine Crépé-Renaudin}
\author[4]{Louis D'Eramo}
\author[7]{Cristinel Diaconu}
\author[2]{Giulio Dujany}
\author[3]{Federico Ferri}
\author[6]{Marie-Hélène Genest}
\author[8]{St\'ephane Lavignac}
\author[9]{Jessica Lev\^eque}
\author[10]{Cyrille Marquet}
\author[11]{Anselmo Meregaglia}
\author[4]{Stephane Monteil}
\author[1]{Carlos Muñoz Camacho}
\author[3]{Louis Portales}
\author[12]{Philippe Schwemling}
\author[6]{Christopher Smith}
\author[4]{Ana~M.~Teixeira}
\author[13]{Michael Winn}

\affil[1]{\footnotesize Universit\'e Paris Saclay, CNRS/IN2P3, IJCLab}
\affil[2]{\footnotesize Universit\'e de Strasbourg, CNRS/IN2P3, IPHC}
\affil[3]{\footnotesize Universit\'e Paris Saclay, CEA/IRFU, DPhP}
\affil[4]{\footnotesize Universit\'e Clermont Auvergne, CNRS/IN2P3, LPCA}
\affil[5]{\footnotesize Universit\'e Claude Bernard Lyon, CNRS/IN2P3, IP2I}
\affil[6]{\footnotesize Universit\'e Grenoble Alpes, CNRS/IN2P3, LPSC}
\affil[7]{\footnotesize Aix Marseille Universit\'e, CNRS/IN2P3, CPPM}
\affil[8]{\footnotesize Universit\'e Paris Saclay, CNRS/INP, CEA/DRF, IPhT}
\affil[9]{\footnotesize Universit\'e Savoie Mont Blanc, CNRS/IN2P3, LAPP}
\affil[10]{\footnotesize Institut Polytechnique de Paris, CNRS/INP, CPHT}
\affil[11]{\footnotesize Universit\'e de Bordeaux, CNRS/IN2P3, LP2I}
\affil[12]{\footnotesize Université Paris Cité, CEA/IRFU, DPhP}
\affil[13]{\footnotesize Universit\'e Paris Saclay, CEA/IRFU, DPhN}

\begin{document}
\maketitle
 
\thispagestyle{empty} 

\vspace{2.cm}
\begin{abstract}
In view of the European Strategy for Particle Physics process, the French HEP community has organized a national process of collecting written contributions and has pursued a series of workshops culminating with a national symposium held in Paris on January 20-21, 2025 that involved over 280 scientists\footnote{\url{https://indico.in2p3.fr/event/34662/}}. The present document summarises the main conclusions of this bottom-up approach centred on the physics and technology motivations\footnote{A document summarizing the viewpoint of the national funding agencies will be submitted in addition.}.  

\end{abstract}

\tableofcontents
\newpage

\pagenumbering{arabic}
\setcounter{page}{1}
\section{Context}
The present document summarises the priorities of the French particle physics community, which have been expressed within the process of the European Strategy for Particle Physics. The community followed a bottom-up approach to discuss the future experimental scenarios in Europe, with a focus on major facilities\footnote{The list of projects and the corresponding parameters presented in Table~\ref{tab:my_label} reflect the current understanding of the future proposals as discussed during the national symposium \url{https://indico.in2p3.fr/event/34662/}} (Table~\ref{tab:my_label}). The approach is centred on scientific and technical excellence, and does not consider in depth other relevant aspects such as human resources or financial planning. It addresses transverse aspects relevant for a solid and diverse research programme, and concludes on scenarios for the long term future of particle physics in Europe.

\begin{table}[h]
    \centering
    \small
    \begin{tabular}{cccccc}
        \toprule    
        Facility & Type & Collision energy & Inst. luminosity [cm$^{-2}$s$^{-1}$]& 
        Timeline
        \\
        \midrule
        LHC         & $pp,pA, AA$  &  $\le 14$ TeV & $2\times 10^{34}$ & 2010--2026\\
        HL-LHC      & $pp,pA, AA$  & $\le 14$ TeV & $7.5\times 10^{34}$ & 2030--2041\\
        Belle II    & $e^+e^-$ & 10.58 GeV & $6\times 10^{35}$ & 2018--2042\\
        \hdashline
        Hyper-K     & $\nu_\mu,\bar\nu_\mu$ & $E_\nu<1$~GeV & & 2027--\\
        DUNE        & $\nu_\mu ,\bar\nu_\mu $ & $E_\nu\sim$1-5~GeV & & 2032--\\
        EIC         & $ep, eA$ & 20--140~GeV & $10^{33}$ -- $10^{34}$ & 2032--\\
        \hdashline
        \FCCee      & $e^+e^-$ & 91--365 GeV & $2\times 10^{36}$ -- $2\times 10^{34}$ & 2045-- \\
        ILC         & $e^+e^-$ & 91--250--550 GeV & $5.4\times 10^{34}$ -- $7.7\times 10^{34}$ & 2043--\\
        CLIC        & $e^+e^-$ & 0.4--3 TeV & $2\times 10^{34}$ -- $6\times 10^{34}$ & 2042--\\
        LEP3        & $e^+e^-$ & 91--240 GeV & $5\times 10^{35}$ -- $1\times 10^{34}$ & 2048--\\
        LHeC        & $ep, eA$ & 1.3 TeV & $2\times 10^{34}$ & 2044--\\
        \FCChh      & $pp,pA,AA$ & 100 TeV  & $3\times 10^{35}$ & >2070\\
        HE-LHC      & $pp,pA,AA$ & 25 TeV  & $2\times 10^{35}$ & 2048--\\
        Muon collider & $\mu^+\mu^-$ & 3--14 TeV  & $2\times 10^{34}$ -- $4\times 10^{35}$ & >2055\\
        \bottomrule   
    \end{tabular}
    \caption{\small Overview of present, approved and proposed facilities discussed in this document.
    }
    \label{tab:my_label}
\end{table}

\section{Physics Motivation}
\label{sec:physics}
The physics motivation is discussed along four thematic axes related to the particle physics research. The larger landscape, emphasising the scientific interest and related facilities relevant in nuclear physics and astrophysics, is addressed in a dedicated section. 

\subsection{The energy frontier}
The past decades have witnessed the success of the Standard Model (SM) in describing particle physics phenomena. Other than the discovery of the Higgs boson, numerous other measurements as well as further relations between observables have strengthened the SM description of strong and electroweak interactions, and its intrinsic mechanism of spontaneous symmetry breaking. 
Nevertheless, new physics (NP) beyond the SM is
clearly needed, to address experimental observations that are not 
explained 
within the SM, such as the neutrino mass generation, the baryon asymmetry of the Universe, and dark matter (DM), as well as to resolve several theoretical shortcomings of the SM itself, including the origin of the electroweak (EW) symmetry breaking or the hierarchy problem. 
In spite of a large number of searches and precise measurements, 
no NP has so far been discovered. 
The path ahead is two-fold: discovery of new states, and tests of the SM through precision and consistency.

In this context, the Higgs/scalar sector is commonly seen as a potential portal for NP signals. Currently, several Higgs-boson couplings have been determined to a 5-10\% precision, while its
mass is known to the per mille level. Determining the Higgs boson self-coupling $g_{HHH}$ (and hence the shape of
the SM scalar potential) will constitute the ultimate test of the SM’s mechanism of EW symmetry breaking,
further shedding light on the stability of the vacuum, and on the possible presence of NP. Precise
determinations of scalar couplings to other fields (and relations between self-couplings) are also sensitive probes of extended scalar sectors, which are a feature of many well-motivated NP models (multi-Higgs doublet models, CP violating scenarios, compositeness, supersymmetry...). 

The full exploitation of the HL-LHC data
should allow for a determination of $g_{HHH}$ with an accuracy better than 50\%. Any indirect sensitivity to $g_{HHH}$
is not only limited by the statistical and systematic experimental uncertainties in the observables, but also
by the available precision on other SM EW and strong parameters, and finally by theoretical uncertainties.
Historically, EW precision tests have played an instrumental role in constructing and testing the SM. The
HL-LHC is expected to improve upon many EW precision measurements (and carry out new ones). Other
than indirectly suggesting the presence of NP (and possibly its scale), precision measurements at the
$Z-$pole and at the $WW$-threshold are critical to our capacity to precisely determine Higgs and EW
couplings. In view of its uniquely large coupling to the Higgs, precise determinations of the top quark
mass, couplings and production cross sections constitute 
essential
tests of the EWSB mechanism, with
the potential to reveal new deviations.

The HL-LHC will be the last high-energy hadron collider for several decades, and the only means
of directly searching for heavy NP states in the near future. Exploiting its full physics potential through the
realisation of the planned detector upgrades – Higgs-boson properties, top-quark physics, QCD and EW
studies, NP searches (including DM and long-lived particles) – is of the utmost importance.
The full
exploitation of the (HL-)LHC legacy requires ensuring that the data (and results) are preserved, and can
be (re)used and (re)interpreted by the whole community. Effort should be put in enforcing systematic
guidelines allowing to do so, and in ensuring that publicly released material  follow the FAIR
principles (Findable, Accessible, Interoperable, and Reusable data and data products).

As recommended by the 
previous European strategy   process, 
the precise measurement of the Higgs boson properties and in particular its couplings is a priority. In
addition, studies of the top quark (especially its
mass) are also extremely important.
With this in mind, the following facilities have been considered based on their R\&D and design
maturity: the \FCCee project which is a 91 km-long $e^+e^-$ collider, an $e^+e^-$ linear collider facility (LCF) à la
CLIC or ILC, both of which could be hosted at CERN. With such machines, the precision on the Higgs
boson couplings can be typically improved by a factor 10 compared to the ones which will be obtained at
the HL-LHC. The \FCCee provides larger statistics compared to an LCF (typically a factor 5-10) thanks to a
higher planned luminosity and multiple interaction points which can operate at the same time. The energy in the
centre of mass ranges from 90 GeV to 365 GeV, i.e. from the $Z-$pole to the $t\bar t$ threshold which allows
for a theoretically clean measurement of the top quark mass. With data collected at $\sqrt{s}=125$~GeV, it could also potentially
probe the electron Yukawa coupling. In addition, on a longer term, such a tunnel could host a high-energy
proton-proton collider (at 100 TeV). An LCF presents the advantage of being able to probe energies up to
500-1000 GeV (with an upgraded version and new technology) which allows direct di-Higgs production
and hence to measure the
Higgs-boson trilinear coupling. The possibility of doing high precision Higgs boson measurements
(especially probing the trilinear coupling) with a high energy proton-proton collider has also been evoked,
even with a relatively low energy machine (27 TeV) with a precision comparable to an LCF. The
community also acknowledges the importance of an 
$ep$ collider programme to ensure the success of a
potential high-energy proton-proton program. The importance of building the next machine at CERN in a
timely fashion was stressed, in order to maintain the expertise and attractiveness of the field for future
physicists.

In addition to the short-term need for a Higgs factory, the community agrees upon the necessity to
engage on longer term developments. It has been stressed that a muon collider able to reach a
centre-of-mass energy $\mathcal{O}(3-14)$~TeV
is one of the most promising projects 
with a physics scope ranging
from Higgs boson precision measurements, to probing the longitudinal vector boson scattering, with a strong 
potential to explore new energy scales, possibly leading to NP discoveries.
All the limitations for such a
project are technical, one of the dominant ones being muon cooling. 
The (growing) participation from French groups is now organised within the international muon collider collaboration.

The effort towards
 precision measurements can be complemented by non-collider approaches such as the measurements of
 $\sin2\theta_W$ at different energy scales, or (new) experimental setups to achieve a relative uncertainty on $\alpha_\text{QED}$ at
 the level of $10^{-11}$. Among these, PAX at ELENA opens a new avenue for strong-field QED tests with exotic atoms,
 and for high-precision QED theory tests using accurate spectroscopy of highly-charged ions. 

\vspace{0.2cm}
\textit{In summary, exploiting the full physics potential of the HL-LHC is of the utmost importance, in particular in view of its unique potential to unravel the present energy frontier and the scalar sector of the SM.  The \FCCee programme, owing to its high integrated luminosity delivered in a short amount of time, is the ideal programme for a significant and timely progress in the understanding of the Higgs boson properties and of EW symmetry breaking. Moreover, \FCCee provides a large amount of data at the $Z$ pole (O($10^{12}$) bosons)  allowing for a large variety of stringent electroweak tests of the SM, and possibly paves the way to a future hadron collider. If such an ambitious project is not deemed feasible, an LCF at CERN (operating up to at least 500 GeV) is a noteworthy
fallback approach. In particular, its high energy programme directly addresses  the Higgs boson trilinear coupling and the top-quark properties.
}

\subsection{The intensity frontier}
Research at the intensity frontier allows the detection and
interpretation of signs of NP. If new particles are discovered directly, indirect tests are needed to study their structure and couplings, and if not, experiments at the intensity frontier can explore higher scales and smaller couplings. The community supports an experimental strategy based on two complementary pillars: experiments with a wide physics program, and smaller dedicated experiments, as detailed below.

\subsubsection{Experiments with a wide physics program}

The major players in flavour physics in the upcoming years will be LHCb and Belle~II: the former exploiting the
unprecedented number of $b$ hadrons produced at the LHC, the latter
profiting from the clean $e^+e^-$ environment of SuperKEKB and a nearly hermetic detector. These two experiments will
improve our understanding of the flavour picture, and probe NP
in a complementary way to that of high precision measurements of a plethora of observables, part of a well-recognized physics program. French physicists have strongly contributed to flavour physics in the past and are now playing a major role in LHCb and Belle~II. Our community emphasises the need to exploit the data from LHCb and Belle~II in the next years. In addition, we  strongly support the proposal of the LHCb collaboration for an “Upgrade~II” of the experiment during the fourth long shutdown of the LHC in 2034, as well as the proposal for a Belle II upgrade around 2032. LHCb Upgrade~II will allow operations at a factor 10 higher instantaneous luminosity, fully exploiting the large luminosity of
the accelerator, to reach unprecedented precision in key measurements, e.g. for the study of CP violation and the unitarity triangle (in particular the angle $\gamma$), $b \to (s,d)\ell^+\ell^-$ and $b \to c \tau \nu$ observables. The recent discoveries of new exotic hadrons have revitalized the field of heavy-flavour spectroscopy. The large statistics that will be obtained with the LHCb Upgrade~II will enable the study of these states in hadronic environments of very high multiplicity, allowing the exploration of their structures dominated by non-perturbative QCD at the femtoscopic scale. The Belle~II upgrade will allow reaching a higher instantaneous luminosity while increasing the detector performance. This will allow world-leading results mainly for final states fully-inclusive or with final states particles with little or no direct signature in the detector, e.g. NP searches in $b \to \ell \nu$, $b \to s\nu \overline{\nu}$, $b \to s \tau^+ \tau^-$, and the most precise determination of the quark-mixing parameters $|V_{ub}|$ and $|V_{cb}|$. It will also provide unique capabilities in $\tau$ physics, including the measurement of its electric dipole moment (EDM) and tests of lepton-flavour violation (LFV) and lepton-flavour universality involving the third generation. An additional possibility is to introduce electron beam polarisation to allow precision measurements of electroweak and $\tau$ lepton parameters.  Belle~II and its upgrade will bring key inputs to develop the technology of the future Higgs factory.

On a longer term,  the best opportunity would come from an $e^+e^-$ collider running at the $Z$ pole accumulating several $10^{12}$ $Z$ events. This will constitute an ideal laboratory for flavour physics: a large sample of all $b$-hadrons, $c$ hadrons and  $\tau$ leptons produced in an clean $e^+e^-$ environment with a large boost.  Direct searches for heavy neutral leptons (as right-handed neutrinos)
will also be made possible. A further run at the $W^+W^-$ threshold collecting several $10^9$ $W^+W^-$ would allow direct access to several parameters of the Cabibbo–Kobayashi–Maskawa (CKM) matrix. To fully exploit these opportunities at least one suitable detector with excellent vertexing, good momentum reconstruction down to low momentum, good particle identification over the full kinematic range, good energy and direction resolution for neutrals and good $K_S^0$ and $\Lambda$ reconstruction efficiency would be needed. These conditions would be met at the \FCCee, the first phase of the FCC project. 

\subsubsection{Experiments designed for specific measurements or searches}
French physicists are involved in experiments dedicated to the search for permanent EDM, which are CP-violating observables predicted to be tiny in the SM, but large in many NP scenarios, in particular those aiming at a viable EW baryogengesis. The present experimental sensitivities for various systems (e.g. neutron, electron) probe scenarios of NP up to energy scales around 100 TeV. In particular, the neutron EDM offers the most sensitive probe for the $\theta_{QCD}$ term. Its search is led by the n2EDM experiment at the Paul Scherrer Institute, expected to reach the $10^{-28}$~ecm region in its second phase in 2030.
Forbidden in the SM, LFV could also be within reach in some NP scenarios. In this context, French physicists are involved in the COMET experiment (COherent Muon to
Electron Transition) at JPARC. It will improve the current single event sensitivity on $\mu \to e$ conversion by two (four) orders of magnitude in a first (second) phase. This will have a strong impact on models predicting LFV and is complementary to the insights from other experiments, e.g., MEGII, Mu3e, as well as LHCb and Belle~II. French physicists have also expressed an interest in the PIONEER experiment aiming at the precise measurements of the ratio of pion leptonic decays, which would provide one of the most precise tests of lepton flavour universality. Finally, in a different context, unique tests can be performed using low energy antiprotons, like the Weak Equivalence principle at the Gbar experiment.

Precision experiments not only constrain the existence of new heavy states with sizeable couplings to known particles, but also that of new light states with very suppressed interactions with normal matter. Those could directly make up the elusive dark matter, or more generally be part of a richer dark sector. An area of intense activity concerns a specific Dark Matter candidate, the axion, whose existence is motivated by the strong CP problem. Many experiments are currently running or planned in the near future, two of which with French involvements, GRAHAL and MADMAX. Both plan to build haloscopes with unprecedented and complementary sensitivities in the search for axion DM, using the new hybrid magnet at LNCMI in Grenoble for GrAHal, and a new dielectric design for the latter.

Finally, experiments at the boundary between particle and nuclear physics can also provide precise tests of the SM or determine some of its key parameters. For the former, nuclear beta decays provide probes for new sources of CP-violation, complementary to EDM searches, and of hypothetical new particles and interactions, while baryon-number violating dinucleon transitions can search for NP up to very high scales. For the latter, superallowed Fermi transitions currently provide the most precise determination of the CKM matrix element $V_{ud}$ (which will also be the target of the PIONEER experiment in its second phase). In view of these examples, the synergy with NuPECC should thus be strongly pursued. French physicists already play a leading role in these studies; in the next decades the radioactive ion beam facility DESIR at GANIL will be uniquely suited for conducting experiments aimed at precision tests of the SM.

\vspace{0.2cm}
\textit{
In summary, the experimental strategy in flavour physics at the intensity frontier is based on two complementary pillars: 
large facilities with a wide physics program and smaller experiments dedicated to specific measurements or searches. A strong support is expressed for the proposed upgrades of LHCb and Belle II. In the long term, the best opportunity  is provided by  an $e^+e^-$ collider accumulating important $\mathcal{O}(10^{12})$ samples of $Z$ and $\mathcal{O}(10^9)$ $W^+W^-$ events. 
In addition, 
dedicated experiments 
have an important
potential for high-impact discoveries, such as long-lived dark particles, as well as for searches for new physics based on observables suppressed or forbidden in the SM, such as EDMs, LFV processes, proton decay, and neutrino-less double beta decay.
}

\subsection{Neutrino Physics}
Neutrinos remain 
the least known elementary particles:
we do not know the absolute value of their masses,
whether their ordering is normal or inverted, whether the charge-parity (CP) symmetry is violated in the lepton sector,
and whether neutrinos are Dirac or Majorana fermions.
Neutrino experiments aim to answer these questions, which are of fundamental
relevance for particle physics,
astrophysics and cosmology.
In particular, the precise study of oscillations with long baselines (LBL) is a privileged tool to investigate standard and non-standard neutrino properties.  

\subsubsection{Long baseline oscillation experiments and beyond}
The advent of LBL experiments pushed the study of neutrino oscillations into the precision era. The availability of controlled sources of neutrinos and antineutrinos, combined with precise flux and cross section measurements before oscillation at near detectors, has allowed the present generation of experiments (T2K and NOVA) to provide accurate measurements of neutrino oscillations, first hints of CP-symmetry violation and new tests of the PMNS oscillation paradigm.

The next generation experiments, Hyper-Kamiokande (HK) in Japan and the Deep Underground Neutrino Experiment (DUNE) in the US, will provide unprecedented sensitivity to CP-violation, determine the mass ordering, and precisely measure oscillation parameters.
HK will exploit a new water Cherenkov detector, and the recently upgraded neutrino beam and T2K near detectors of T2K, whereas DUNE will rely on liquid argon time projection chambers, and a new neutrino beam and near detector complex at Fermilab.
These experiments will enable a deeper understanding of neutrino properties and provide powerful tests of New Physics scenarios. Moreover, the far detectors of LBL experiments have the best sensitivity to proton decay and can detect neutrinos from the Sun, from supernovae and from other astrophysical sources.

In the longer term, the ESSnuSB/ESSnuSB+ projects, funded by Europe, investigate a unique possibility 
of creating a LBL experiment with unprecedented beam power in Europe, also fostering R\&D for future neutrino factories and monitored/tagged neutrino beams.

Beyond LBL, French physicists are engaged in an ambitious and diverse neutrino physics program.
JUNO and ORCA, studying oscillations at reactors and from atmospheric neutrinos, will provide improved precision in the oscillation parameters and strongly contribute to the determination of mass ordering.  KM3NeT is a European-hosted project aimed at becoming a major future player in the blooming domain of multi-messenger astronomy. Europe is also hosting world leading bolometric experiments for the search of neutrinoless double-beta decay: CUORE, and its successor CUPID. Finally, the French neutrino community is engaged in innovative R\&D efforts for low energy detectors that could enable new tests of the Standard Model.

\subsubsection{Role and opportunities for the CERN Neutrino Platform}
The CERN Neutrino Platform is a cornerstone of the European contribution to the worldwide neutrino physics program, supporting the design, testing, and validation of cutting-edge detectors for T2K, HK, and DUNE. It provides invaluable infrastructure for detector assembly, cryogenic tests, and beam line studies, offering facilities unmatched in Europe, thus enabling European institutions to play key roles in global neutrino collaborations. In order for Europe to achieve a critical mass for overseas projects, a coherent European strategy and a centralized hub are needed to share and mutualize resources and expertise. The availability of a fully functional environment at CERN, easily accessible from Europe, is a unique way to enable a strong European participation in experiments overseas. 

Continued support of the Neutrino Platform will be crucial for further detector developments envisaged for DUNE, with the module of opportunity, and for HK, with the second upgrade of the ND280 near detector. A future platform expansion to support detector development for a variety of neutrino experiments (e.g. photo-detection technology for neutrino telescopes, radio-purity studies for low energy solar neutrinos measurements and $0\nu\beta\beta$) would be a major asset to further optimize resources and share expertise.

The precision measurements of neutrino oscillations will ultimately be limited by uncertainties due to nuclear physics effects in hadro-production in the beamline and in neutrino-nucleus interactions. Measurements of hadro-production (in NA61/SHINE) and of hadron and electron scattering on nuclei are of paramount importance to ensure the accuracy of neutrino oscillation measurements. With its proton, pion and electron beams, CERN is a unique environment for such crucial measurements. Promoting CERN as the worldwide platform for such a program is a compelling opportunity to build a unique and distinct role for the European  community in the neutrino  physics field.

A neutrino beam at CERN, as currently discussed at SBN@PBC (Short Baseline Neutrinos at Physics Beyond Colliders), would provide essential insights into neutrino interactions, offering precision measurements and an ideal platform to test new detector concepts. Initiatives like ENUBET, which aims to deliver a monitored neutrino beam for precision cross-section measurements and NuTAG, focused on detecting neutrino interactions with unprecedented detail, demonstrate the potential of these experimental efforts to boost the physics return of LBL projects. Such a beam could be directed at existing detectors, such as the ProtoDUNE modules and Water Cherenkov Test Experiment (WCTE), adding value to previous investments. 
These projects will enable new technological solutions for the long term future of the LBL domain.

The continuation and strengthening of the Neutrino Platform will be crucial to maintain Europe's critical role in neutrino physics. As a centralized resource, the platform exemplifies the benefits of collaboration and shared expertise.

\subsubsection{CERN as a hub for analysis and for theory}
Theory is a crucial part of the general effort by the neutrino community to understand the fundamental
properties of neutrinos and identify the physics at the origin of their masses. Sustained activity in neutrino theory is mandatory to fully exploit
and interpret the experimental results, and to guide experimental searches.

CERN serves as a hub for collaboration and exchange in neutrino physics, bringing together experimentalists and theorists from across Europe and beyond. Initiatives such as the annual CERN Neutrino Platform Pheno Week and the Physics Beyond Collider efforts foster dialogue between the communities, enabling the development of theoretical models that inform experimental designs and interpretations. CERN also provides a powerful infrastructure for shared software platforms, Monte Carlo simulations, and computational resources, facilitating consistent data analysis across multiple experiments. 

The advent of HK and DUNE raises the challenge of comparing and combining their results. Due to strongly correlated systematic uncertainties, close collaboration will be essential to ensure a consistent interpretation. CERN can play a key role as a promoter and host of such joint efforts, leveraging the strong involvement of the European community in both experiments, benefiting both the LBL science goals and the visibility of European neutrino research.

\vspace{0.2cm}
\textit{In summary, the variety and importance of the neutrino physics program in Europe is a great asset for the community and is expected to play a major role in the overall physics landscape in the next decade. It is therefore important that the next collider project at CERN be affordable in a way that preserves the support to neutrino physics.
In particular, a centralised European approach to neutrino physics at CERN, relying on the Neutrino Platform infrastructure and using CERN as a hub for analysis and theory, is necessary to ensure that the European physics community remains at the forefront
of future scientific discoveries in the neutrino field.}

\subsection{Strong Interactions}
The study of the strong interactions is essential for the understanding of the universe.
 Complementary and precise measurements are provided from many different areas: from accelerator-based facilities and lattice QCD, to atomic physics and gravitational waves. 
 Moreover, strong interactions effects are present in observables and often limit the precision of other measurements (e.g. electroweak parameters) or the searches for new physics.

High-energy collisions of nuclei offer unique tests of the high-temperature thermodynamics of the SM in laboratory-based experiments, at energy densities where partonic degrees of freedom dominate the equilibration processes,
and determine the properties of the high temperature phase of 
QCD, the quark–gluon plasma (QGP).  
LHC experiments will pursue the exploration of the QCD phase diagram
with better precision, new observables, and exploiting a range of initial states. Run 5 will be crucial in understanding the evolution of collectivity from small to large systems, in collisions of protons and nuclei of increasing size. 

The LHC Run 5 will include the full exploitation of a fixed-target programme led by the LHCb collaboration, with precise measurements of heavy-flavour and quarkonium production in $pp$, $pA$ and $AA$ collisions over a broad rapidity range. Moreover, instrumentation up to high rapidities will enable unique measurements for hadron structure. 
The LHCb-U2 project 
offers the opportunity to study heavy-ion collisions up to the most central collisions with the LHCb detector.  
The emphasis of the programme will be on heavy-flavour and muonic final states, but will not be limited to these, since the acceptance, resolutions, trigger and particle-identification capabilities offer measurement opportunities for a variety of/ final states including photon-induced reactions. 

The scientific program with heavy ions will address two different and complementary physics quests. First, the origin of the collective-like phenomena observed in very light systems such as high-multiplicity $pp$ collisions can be studied in detail with LHCb, through measurements of intermediate systems between $pp$ and Pb-Pb collisions, involving protons as well as Oxygen, Xenon and possibly other nuclei. 
In addition, LHCb is a unique experiment for proton-nucleus and nucleus-nucleus fixed-target physics at the LHC, at centre-of-mass energies close to those of RHIC and under very different experimental conditions.  

The centre-of mass energy and rapidity coverage, in collider and fixed-target modes, and including ultra-peripheral and polarised fixed-target collisions, will play a decisive role in the characterisation of the QCD equation of state, and the mechanisms of deconfinement, thermalisation and hadronisation. 
As a next-generation dedicated heavy-ion detector at the LHC, ALICE~3 is designed to span a broad and ambitious physics program involving both precision and exploratory measurements, from pp to Pb-Pb collisions, with the final goal of improving our understanding of the mechanisms responsible for the formation and the behaviour of complex hadronic systems and phases, and how the properties of these systems connect to the fundamental parameters of QCD. 

Lepton-hadron collisions at the GeV scale and above offer the unique possibility of probing the structure of hadrons at the partonic level with the well-understood electromagnetic interaction 
in the initial state. The EIC will collide polarized electrons with a large variety of ions and deliver large luminosity,
allowing the study of the structure of matter at very small momentum fraction $x$ and investigating the prominent role of gluons in QCD. In conjunction with measurements at the HL-LHC, EIC can shed light on gluon saturation effects, a new state of matter predicted by QCD but so far not unambiguously observed.
It will enable the multidimensional tomography of nucleons. 
In the QGP research domain, 
a fruitful synergy relates the EIC and the LHC heavy-ion programmes, including proton-ion and photon-ion reactions.  
Therefore, the EIC science program plays a strategic role in the landscape of nuclear and particle physics in
Europe and worldwide: from low-energy nuclear physics to high-energy particle physics, including
QCD theory and phenomenology.

The FCC, in its electron-positron and hadron-hadron running modes, will be an outstanding facility for new and precise QCD
measurements. The \FCCee will produce extraordinarily large samples of electroweak boson decays in hadronic final states. 
High-precision QCD measurements (the coupling constant $\alpha_S$, parton shower and hadronisation studies, hadron spectroscopy) are interesting per se, but also crucial for indirect searches of physics beyond the SM. Collisions of virtual photons, $\gamma^{*}\gamma^{*}\to X$, will allow a new type of high-energy QCD measurements to be performed over a wide range of momentum transfers. 

Pb-Pb collisions at the \FCChh 
will produce a deconfined state of QCD matter at unprecedented energy densities. At these energies, charm quarks become an active degree of freedom. 
The large collision energies and integrated luminosities will allow new types of ``tomographic'' plasma probes. 
The high density and high energy of the photon field in ultraperipheral heavy-ion collisions will provide novel searches for new physics  using photon-photon collisions. Very low gluon fractional momenta (down to $x\approx 10^{-7}$) in the nuclear parton densities will be explored with perturbative probes in proton-lead collisions. 
Furthermore, \FCChh will study the QGP at unprecedented energy densities. 

In the perspective of the \FCChh, a high-energy electron-proton collider, the LHeC, would be able to provide the parton distribution measurements required for the exploitation of the \FCChh data and has a unique potential for gluon saturation physics. Opportunities to explore the high-net-baryon density region of the QCD phase diagram  at fixed-target energies with dedicated experiments at CERN
 complementary to other facilities should be explored in parallel to the next CERN flagship project after the LHC. This has to be put in perspective with  gravitational waves projects. 

Understanding measurements in lepton-hadron, hadron-hadron collisions and low-energy tests of the standard model requires a sustained support to developments in the theory of strong interactions. These developments follow a wide range of approaches, from formal theory up to the full description of exclusive events through event generators, and prominently include methods addressing the non-perturbative regime of QCD. 
A current highlight is the theoretical determination of $(g-2)_{\mu}$, where France plays a key role in the determination of the QCD-related vacuum polarisation and its uncertainties.

\vspace{0.2cm}
\textit{In summary, the priorities for strong interaction physics are:
the physics program with ions during the full lifetime of the HL-LHC for the understanding of QCD processes, 
the  LHCb upgrade for the Run 5 and ALICE 3 experimental program with their high potential for exploring new territories in heavy ions physics, 
the strong complementarity between the heavy-ion program at LHC Run 5 and the EIC project in the US  and the importance of both projects on the European future roadmap,
the next generations facilities \FCCee and \FCChh with  a very broad, different and complementary QCD physics programs.
}

\subsection{Particle physics and the larger landscape}

Besides the flagship projects such as LHC, the particle physics ecosystem includes low-energy infrastructures at CERN, like the SPS for fixed-target experiments, and unique facilities like ELENA, as well as infrastructures elsewhere in Europe, such as the Modane and Gran Sasso underground laboratories, the Mainz and PSI accelerators for low-energy physics (with high-intensity electron and proton beams), the Ganil DESIR facility, and the DESY laboratory.  These infrastructures are the core of a diversified and fruitful European particle physics strategy and must be supported.  

 Besides the experimental lines of research and experiments beyond colliders already mentioned in Section~\ref{sec:physics}, direct searches for WIMPs (e.g. DarkSide,  XENON, or TESSERACT) should also continue to be pursued, as they offer a unique opportunity to address open questions on DM. With the growing size of these
 experiments, creating a new European network fostering the collaboration of
 the involved institutions on technical, scientific, organisational and funding aspects could become interesting. 
 
The community also notes the growing importance of gravitational-wave (GW) experiments whose physics program has deep connections with particle physics. CERN holds a unique technological expertise of critical value for the design and construction of ground-based GW telescopes. %
 An increased role of CERN in this domain would be a great asset for the European community and 
would bring to the CERN portfolio an additional high-profile physics case 
with a large and ensured physics return.  The third-generation projects such as the Einstein Telescope could also bring new cutting-edge technologies to CERN, the development of which will benefit future particle physics projects. 

The strong connections between particles physics and atomic physics, nuclear physics and astrophysics calls for enhanced coordination, like in the JENAS initiative: this is especially important for projects at the interface between these domains and to ensure the most effective cross-fertilization between them.

Europe should also continue to invest on lines of physics research with direct societal applications, like imaging techniques for biomedical applications, as well as on interdisciplinary efforts which could serve the study of our natural environment, like the 'Laboratoire sous-marin Provence Méditerranée'.

\vspace{0,2 cm}
\textit{In summary, the present status of investigations into the SM
requires a
diversified approach:
 in particular, it is important to keep investing in smaller dedicated experiments with
potential high-impact discoveries. 
To this aim, the diversity of the CERN and of the larger European particle physics research ecosystem must be preserved 
as a strategic source of scientific and technological innovation.
}

\section{Further relevant considerations}

\subsection{Theory}
Theory is a crucial part of the general effort 
to understand the fundamental laws of nature and to identify the NP required to address the shortcomings of the SM. In the quest for NP, precision tests of the SM play a crucial role, especially in view of the preparation of future leptonic Higgs factories (also offering the Tera-Z run): theory uncertainties must be strongly reduced to be on par with the expected experimental accuracy. Likewise, precision calculations of flavour observables and of neutrino cross sections are also very important. The assessment of the theory errors requires massive developments of both analytical and numerical integration techniques: this includes EW and QCD higher order contributions (2 to 4-loops), as well as innovative  multi-purpose event generators; such developments can strongly benefit from Machine Learning (ML) methods. Lattice QCD is also critical to access new observables (in particular in flavour physics), and is central to the study of QCD collective effects. 
At the crossroads of theoretical and experimental particle physics ultimately 
lies the interpretation of data, which can be pursued through model-independent characterisations of NP (via an effective approach) and dedicated studies of specific NP models. These can be complemented by new approaches, for instance relying on proto-model builders. Model building approaches offer complete high-energy theoretical constructions; tests of the latter can then inspire dedicated experimental searches. The reconstruction of the NP model will call upon input from all experimental frontiers: high-energy, flavour and EW precision, neutrino physics, QCD, dark matter searches and cosmology.

Any strategy on the future of experimental particle physics must thus be accompanied by a vision concerning the support that must be provided to theoretical particle physics, and by efforts to further foster European initiatives promoting theory-experiment collaborations.

\subsection{Computing, software and data handling}
Computing, software and data handling are essential elements to exploit the collected HEP samples and achieve outstanding physics results.
The HEP data volume is continually increasing, already reaching the exascale. The increased granularity and timing precision of upgraded and future detectors, along with larger samples of real and simulated data, will further enhance this and HEP will stay one of the largest provider of scientific data. Moreover, data should be preserved at long term and the Open Science tools, methods and policies need to be deployed to ensure data availability
and to maximize the scientific return.
At the same time, the industry and academia develop, in an ever rapid pace,  technologies 
 such as Artificial Intelligence and heterogeneous hardware architectures opening both opportunities and complexities for HEP, while facing simultaneously compelling demands to reduce their environmental impact. 

In this context, it is crucial 
to keep a dedicated infrastructure at the core of our computing systems to provide reliable and efficient resources, adapted to our needs but also sufficiently agile and flexible to fully leverage advancements in hardware and data analysis methods. The utilization of external resources, such as HPC and cloud services, must be evaluated meticulously in terms of their advantages and the preservation of technical know-how of our community.
Bringing computing very close to the data acquisition system, with the quality necessary for immediate physics analyses is also a fast-developing field. Highly efficient and energy-saving heterogeneous data intensive 
computing and data distributed systems are a niche where HEP collaborations have a chance to create solutions that will have a broader impact.

Equally important is the continuous development of the software tools needed at all levels: middleware, databases, data transfers, event generation and simulation, data reconstruction and analysis. Strong, dedicated software development teams are needed to maintain, develop, and adapt these tools, making the best use of the new technologies, which is even more challenging in an heterogeneous technical landscape.

Large investments in software and infrastructure, together with
 important and well coordinated R\&D efforts are mandatory, including trans-disciplinary initiatives
such as the HEP Software Foundation, or the joint ECFA-NuPECC-APPEC working groups.
Open Science projects like ESCAPE related to the construction of EOSC are also place where developments across discipline can be pooled, benefiting all parties involved. 
To facilitate an efficient organization in HEP, models similar to those used for Detector R\&D Collaborations should be considered for--or integrated with--software and computing. Experiments must be designed considering the interplay between instrumentation, data analysis and computing requirements.

Addressing those challenges requires the commitment of dedicated computing teams and of the whole HEP community with the need to enhance its computing expertise. Providing training to reach that goal and hiring young scientists with strong IT skills are essential. Promoting new technologies, ambitious goals and attractive research environment is crucial amid competition from other fields and private companies.

\subsection{R\&D activties}

\subsubsection{Detector R\&D} 

A strong R\&D development program is vital to prepare the next generation of experiments. 
It also plays an important role in attracting talented students and engineers to our field, and should offer them opportunities for bright careers.  

In the past two years, CERN and ECFA have started to structure detector and instrumentation R\&D through the DRD collaborations. The French community is very supportive and largely involved with a dozen of laboratories participating in the different DRD areas.  

Most activities target the strategic collider projects identified by the previous ESPP.  They concern the upgrade programs of existing experiments (Alice-3, LHCb-II, Belle-3, ATLAS, CMS), the preparation of the EPIC detector at EIC to be completed around 2035, or the longer term $e^+e^-$ collider projects. An initial R\&D period of 3-4 years should deliver technical solutions for the earlier projects. The expertise gathered in this first phase will be pushed further to match technical requirements for a \FCCee completion around 2045. Continuous R\&D activity is required to follow the fast technological evolutions. 

Mainstream areas being explored towards \FCCee are reflected by EoI contributions to the ESPP for High Granularity calorimetry in an ILD/CLD detector concept, Liquified Noble Gas calorimetry in the Allegro detector concept, and use of Monolithic CMOS sensors in vertex and tracking systems. 

Intense R\&D activities are also targeting the domain of Neutrino, Dark Matter and rare decays physics, with the specific challenges of aiming for extremely low thresholds, the reduction of radioactive backgrounds and the mastering of complex nuclear effects for proper detector calibration.

R\&D activities will need to ramp up during the first short term phase, requiring a substantial increase in the allocated resources. Projects need to be supported  in the area of sustainability and towards the minimisation of the detector environmental impact. Sufficient access to test-beam facilities will be important, in a context where running machines and beam lines will be less accessible than today, with the planned long shutdown of the CERN accelerator complex.

\subsubsection{Accelerator R\&D} 

Accelerator R\&D has an intrinsic value and should be pursued along a well structured and balanced program in order to maintain Europe's leadership in this area. 
Research and development in energy recovery, through projects like PERLE and the LHeC, is a key focus, particularly in the context of the sustainability requirements discussed below.
High-field dipole technology remains a crucial area of R\&D and should be actively pursued to maximize the energy reach of future colliders. Similarly, the SCRF technology is important for the development of future particle accelerators, as it enables higher acceleration gradients and improved energy efficiency compared to conventional technologies.
Innovative developments in accelerator technology, such as plasma acceleration (e.g., AWAKE), hold great potential. These innovations could enable more compact accelerators with applications ranging from particle physics to broader societal benefits.
A muon collider presents significant challenges, including the need for advancements in high-temperature superconductors (HTS). Its realization requires long-term R\&D beyond the scope of the present discussion.

\subsection{Sustainability}

Sustainability considerations in HEP are pivotal, to respect the planetary boundaries, to comply with the rapidly evolving regulation, and to align with the global effort demanded on society. The HEP community should lead by example by addressing these issues from the earliest stages of future projects, thereby increasing their acceptance by civil society and strengthening the staff's engagement.

Maximizing the scientific return of HEP projects must be balanced against costs, efficiency, and environmental footprint, with the ambition
to reach an overall environmental impact
compatible with global decarbonization goals. The environmental factors, including life cycle analysis and socio-economic impact studies, must be
among the core evaluation criteria  
to be scrutinized in-depth.  
The quantification and mitigation of environmental footprints across laboratories, institutes, and collaborations must be generalized. Sustainable procurement practices and building construction, energy mix considerations and the promotion of new research practices also require special attention. The new infrastructures should be eco-designed, prioritising the use of low-carbon material, reducing negative impacts on biodiversity and ideally offering it a haven. The minimization of the impacts should be seen as an opportunity to explore new R\&D paths rather than a constraint limiting the scientific goals. The integration of a life cycle assessment in the future accelerator project proposals is considered as a positive first step. 

To go beyond aspirations, significant investments in R\&D for sustainable detector technologies, energy-efficient computing powered on low-carbon electricity, and innovative acceleration methods are crucial. Priorities encompass phasing out high-GWP gases for cooling or particle detection, and energy efficiency in operations. France’s expertise in areas such as energy recovery accelerators and klystron efficiency offers opportunities to contribute to an European leadership in sustainable technologies. Committing to a sustainable trajectory is also an opportunity to bring new expertise on cutting-edge technologies in HEP.

Finally, communicating the efforts to minimize the environmental footprint to the rest of the scientific community, decision-makers and the general public is crucial for greater societal acceptance, particularly for large-scale projects. France, as host state, has a leading role to play in this respect. The communication strategy would benefit from the definition of transparent and unbiased key quantitative indicators and a clear carbon footprint reduction trajectory, with long-term planning.

\subsection{The role of Early Career Researchers }

The Early Career Researchers (ECR) should continue to play a central role in the field of high-energy physics. By the start of next-generation particle collider operation, the ECR of today will in all likelihood be the main drivers and leaders of the projects, and all possible means should be taken now so that they can succeed in these roles. An important aspect in this direction is an effective transfer of expertise.
This includes already, the hands-on experience acquired through the analysis of physics data from the ongoing experiments. 
The 
expertise in electronics, information technology, instrumentation, and mechanics are considered as crucial skills, which should be handed over by involving as much as possible the younger colleagues in the R\&D and design of the future experiments.
A carefully planned guidance should be pursued, especially for students, on how to balance their implication between current experiments and the future-oriented projects. 
The survey of emerging research fields and of new technologies, as well as the training in project management should be encouraged as well.

A clear an prompt decision on the investment plans is requested, in particular on the sequence of the major projects at CERN, in order to improve the confidence in successful career in HEP. 
In this context, the sustainability is considered as a cornerstone of the decision-making process.
Finally, efforts should be made to better communicate on the relevance of the various future 
colliders
to ECRs 
not directly involved in these projects.

\section{Executive Summary: Scenarios for flagship projects in Europe}
The ongoing ESPP process is expected to lead to a clear decision on the preferred next large scale facility at CERN, as a successor of the High-Luminosity LHC.
Indeed, the Standard Model has been verified with a high precision and the knowledge frontiers are constantly overpassed by the intense investigations and precise measurements at the LHC and elsewhere.
Consequently, the particle physics landscape calls for a 
significant step in precision
and the community's  sense of priority
for an $e^+e^-$ collider reached at the former ESPP is still valid.
Together with the specific physics axes described above, transverse and global aspects have been addressed as ingredients of projecting the best possible experimental scenarios for the particle physics in Europe.
Among the consensual views that have been expressed, one can cite the need to determine the energy scale of new physics, through a new body of high-precision data as stated above; 
the need for a flagship project at CERN that would start seamlessly after the completion of HL-LHC; 
the need for ensuring experimental diversity;
the need for sustained R\&D in support of energy-frontier facilities; 
the attention to be paid from an early stage of the future next flagship project to ensure its sustainability while emphasizing its positive societal impact.

\subsection{Preferred option for the next collider at CERN: \texorpdfstring{\FCCee}{FCCee}}

The French community expressed a strong support for the $e^+e^-$ Future Circular Collider project, \FCCee, as the next Flagship facility at CERN. 
The \FCCee project has a compelling physics program, addressing a broad range of physics questions. It will provide great advances in the knowledge of the couplings of the Higgs bosons, the electroweak and strong gauge couplings, prominent electroweak observables ($m_W$, $m_Z$, $m_{t}$ and the corresponding widths), flavour physics (the $b$, $c$, $\tau$ fermions), as well as searches for dark- or light-sector particles. The operation at the $Z$ pole and the related wealth of electroweak Flavour and QCD measurements is considered as a unique opportunity, improving many fundamental measurements by factors 10--100 compared to the legacy from LEP1. 

In a later stage, this option allows for a 100 TeV hadron collider 
re-using the \FCCee tunnel, providing a unique opportunity to explore the energy frontier, and to determine the Higgs-boson trilinear coupling with percent-level precision. Together with a possible electron-hadron collider, this complex would form a major tool for the study of strong interactions.

CERN is the best place worldwide to host a project of \FCCee's size and category, in terms of expertise, connections, international practices, infrastructure and sustainability - including a favourable energy mix availability. In particular, the requirement of sustainability in the design and construction of the facility is considered as a necessity.

\subsection{Fall-back options in case the \FCCee is not feasible} 

This section examines the available scenarios in case the \FCCee is deemed not feasible. Possible reasons include insufficient funding, insufficient societal support, environmental cost, or the confirmed construction of a competitive project outside of Europe and ahead of the European project (for instance ILC in Japan, or CEPC in China). It is not the purpose of this document to define the criteria driving these decisions, but to put forward, in accordance with the ECFA guidelines, a sequence of fall-back projects in case such a decision is made by the appropriate bodies after a careful optimisation of cost, scope and schedule. 

\vspace{0.25cm}
\textit{If the construction of an $e^+e^-$ collider comparable to the \FCCee is not firmly established outside of Europe:}

    In absence of \FCCee, a linear $e^+e^-$ collider facility (LCF) at CERN would be the next best option for a Higgs factory. Somewhat limited statistics at the $HZ$ cross-section peak and a much smaller luminosity at the $Z$-pole are in part compensated by the possibility to reach at least $\sqrt{s}=500$~GeV, allowing a clean observation of the $ e^+e^- \to \nu\nu H$ process, of the $t\bar{t}$ threshold, and providing an improved determination of the Higgs-boson self coupling. Energies of $\sqrt{s}=1$--3~TeV, as enabled by CLIC technology, would significantly improve these measurements and allow detailed studies of vector-boson scattering. 
The LCF program could be complemented by a dedicated, high-luminosity $Z$ factory, possibly re-using existing infrastructure at CERN.

As a last-resort fall-back, LEP3 offers an instantaneous luminosity five times less than \FCCee and an energy range limited to about $\sqrt{s}=240$~GeV. This still matches the purpose of an electroweak, flavour and Higgs factory in line with the 2020 ESPP, but does not allow a complete test of the electroweak theory. LEP3 could be followed by a hadron collider, benefitting of high-field magnets which would be developed at the horizon of the completion of HL-LHC, to address the missing issues on a much longer timescale.

\vspace{0.25cm}
\textit{If the construction of an $e^+e^-$ collider comparable to the \FCCee is firmly established outside of Europe, and ahead of the European project:}

The LCF would provide sufficient scientific complementarity only if it covers the entire energy range between the $t\bar{t}$ production threshold and the TeV scale on a reasonable timespan.

Alternatively, the strategy could shift towards the earlier development of a high-energy $hh/eh$ program, ideally implemented in a new tunnel as in the case of FCC. Given the shorter time available for magnet development, the energy reach would likely be reduced to about $\sqrt{s}=85$~TeV.

If a new tunnel is not feasible, a collider such as the HE-LHC could 
be a fallback alternative with comparable scientific breadth. 
Due to its limited size, the HE-LHC energy reach would be limited to \cal{O}(25 TeV).  However, it would serve as a natural extension of the HL-LHC, reaching similar precision as a TeV-scale LCF. 
Dedicated flavour and HI experiments 
could improve on LHC's respective legacies. 

Both the \FCChh and the HE-LHC should be complemented by an electron-hadron collider such as the LHeC to resolve the uncertainties stemming from the proton and nuclear structure in these uncharted energy regimes. The LHeC also has a rich physics program of its own, with fundamental measurements in the strong and electroweak sectors, and a non negligible potential for NP searches. It could run in the mid 2040's and use improved acceleration techniques based on ERL that will help achieve the sustainability requirements and benefit to future $e^+e^-$ colliders. 

While the fall-back scenarios presented above are clearly sub-optimal compared to the FCC program, the scientific loss is in part compensated by the faster scientific return and increased complementarity offered by a program including $ee$, $pp$ and $ep$ collisions in different regions and on similar timescales.


\vspace{0.3cm}
\textit{ In summary, the huge progress from all areas and experiments in the past five years, in particular from the LHC, sets a solid base for the future of particle physics in Europe. The rapidly evolving scientific landscape demonstrates the relevance of probing the intimate structure of matter on the energy, precision and complexity frontiers, in particular by a full exploitation of the LHC potential. The observed robustness of the Standard Model calls to keep a diversified ecosystem of physics research, including major experiments probing the neutrino- and the dark-sector and small-scale experiments for specific physics cases with potential for high-impact discoveries. 
The important societal and transverse items, in particular sustainability and ECRs career perspectives, should be incorporated as optimization constraints for the future programs. 
A strong support is expressed for a circular electron-positron collider at CERN, since it offers the most complete and attractive physics program after the completion of LHC.  The \FCCee will be at the forefront of the fundamental physics and will stimulate cutting edge technology developments in detectors and data analysis, thereby contributing to Europe's strategic leadership.  
}

\end{document}